\documentclass[fleqn,twoside]{article}
\usepackage{espcrc2}
\usepackage{graphicx}
\usepackage[figuresright]{rotating}

\newcommand{\AmS}{{\protect\the\textfont2
  A\kern-.1667em\lower.5ex\hbox{M}\kern-.125emS}}

\title{The order of the chiral transition in $N_f=2$ QCD\thanks{Contribution based on the talks presented by A.~Di~Giacomo and C.~Pica.}}

\author{
    M. D'Elia\address[GENO]{Dipartimento di Fisica dell'Universit{\`a} di
    Genova and INFN sezione di Genova, Italy},
    A. Di Giacomo\address[PISA]{Dipartimento di Fisica dell'Universit\`a di
    Pisa and INFN sezione di Pisa, Italy},
    C. Pica\addressmark[PISA]
} 

\begin{document}

\begin{abstract}
The order of the chiral transition for $N_f=2$ is an interesting probe of the QCD vacuum.
A strategy is developed to investigate the order of the transition using finite size scaling and its relation to color confinement.
An in-depth numerical investigation has been performed with KS fermions on lattices with $N_t=4$ and \mbox{$N_s=12,16,20,24,32$} and quark masses $am_q$ ranging from 0.01335 to 0.35.
The specific heat and a number of susceptibilities have been measured and compared with the expectation of an $O(4)$ second order and a first order phase transition.
A second order $O(4)$ is excluded, whilst data are consistent with a first order.
\end{abstract}

\maketitle

\begin{figure}[b!]
\includegraphics*[width=\columnwidth]{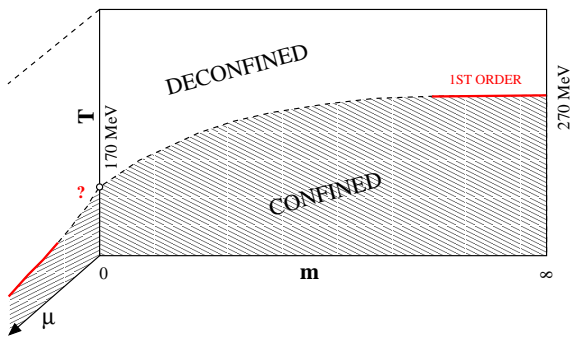}
\vspace{-30pt}
\caption{Schematic phase diagram of $N_f=2$ QCD.}\label{PHDIA}
\end{figure}

QCD with $N_f=2$, is a key system to understand confinement.
The phase diagram for two flavors of mass degenerate quarks is schematically depicted in Fig.~\ref{PHDIA}, where the deconfinement temperature $T_c$ is plotted versus the quark mass and the baryonic chemical potential $\mu$.
At $\mu=0$ the phase transition is well understood at high masses ($m\geq2.5$ $GeV$), where the quarks decouple: as in the quenched limit $m=\infty$ the transition is first order and $\langle L \rangle$, the Polyakov line, is a good order parameter.
\begin{figure*}[t!]
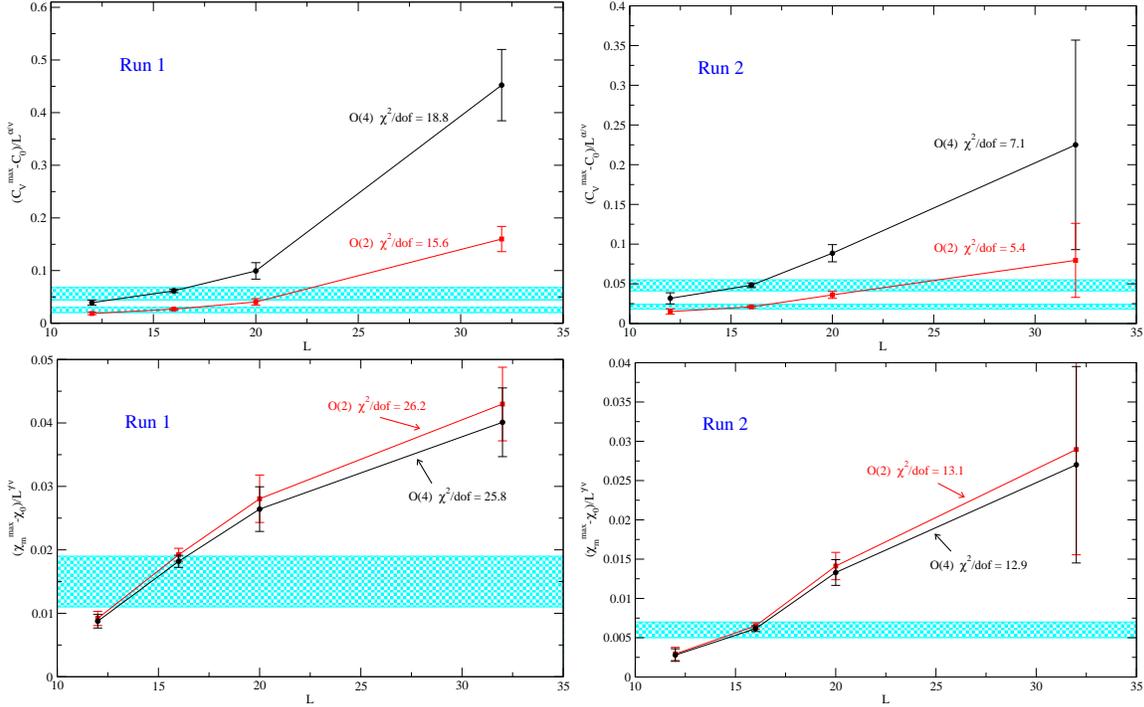

\includegraphics*[width=\columnwidth]{Cv_max_Run1.eps}
\includegraphics*[width=\columnwidth]{Cv_max_Run2.eps}\\
\includegraphics*[width=\columnwidth]{Chi_max_Run1.eps}
\includegraphics*[width=\columnwidth]{Chi_max_Run2.eps}
\vspace{-30pt}
\caption{Specific heat (top) and $\chi_m$ (bottom) peak value for Run1 (left) and for Run2 (right), divided by the appropriate powers of $L_s$ (Eqs.~\ref{CV}-\ref{CHI}) to give a constant. Horizontal stripes indicate the $1\sigma$ confidence region for a fit with a constant value. $\chi^2/dof$ is also shown.}\label{R12}
\end{figure*}
At $m\simeq 0$ a chiral transition exists, where chiral symmetry is restored, and the chiral condensate $\langle\bar\psi\psi\rangle$ is an order parameter.
At some temperature also the axial $U_A(1)$ is expected to be restored, and in fact the topological susceptibility drops to zero around $T_c$~\cite{TOPSUSC}.
In principle at $m\simeq 0$ there are 3 transitions (chiral, axial $U_A(1)$, deconfinement): it is not clear if they coincide, and the question cannot even be asked if there is no independent definition of deconfinenement.
An effective description of the chiral transition can be given in terms of an effective free energy, by usual renormalization group plus $\epsilon$-expansion techniques~\cite{piwi}.
Under the assumption that the scalar and pseudoscalar modes are the relevant critical degrees of freedom, there exist no infrared stable fixed points for $N_f\geq 3$, so that the transition is expected to be first order.
For $N_f=2$ the transition is first order if the anomaly is negligible ($m_{\eta'}\approx 0$) at $T_c$ (the symmetry is $O(4)\times O(2)$); it can be second order with symmetry $O(4)$ if the anomaly survives the chiral transition.
In the first case the transition surface around $m=0$ is first order, and there exists no tricritical point in the $\mu-T$ plane.
If the chiral transition is instead second order, then by general arguments, the surface is a crossover, and a tricritical point is expected in the $\mu-T$ plane (see e.g.~\cite{TRICRI}).

If the latter is the case, the deconfining transition cannot be order-disorder, there is no order parameter for confinement, and a state of a free quark can continuously be transfered below the ``deconfining temperature''.
The transition line in Fig.~\ref{PHDIA} is defined by the maxima of a number of susceptibilities (the specific heat $C_V$, the chiral susceptibilitiy, \ldots) which happen to coincide within errors.
Existing literature on the subject is admittedly not conclusive about the order of the chiral transition~\cite{fuku,colombia,karsch,jlqcd,bernard}, with a diffuse tendency to assume a second order $O(4)$ and crossover.
The problem is fundamental and deserves additional study.

\section{Strategies}

The order and the universality class of the transition can be determined by a finite size scaling analysis of susceptibilities.
The approch is based on the renormalization group, and is valid if the correlation length of the order parameter $\xi$ goes large with respect to the lattice spacing $a$, so that $a/\xi\approx 0$: this is true for second order and for weak first order transition~\cite{FISHER72,BREZIN82}.
A key quantity is the specific heat $C_V$, for which 
\begin{equation}
C_V-C_0 = L_s^{\alpha/\nu}\Phi_C(\tau L_s^{1/\nu}, am_q L_s^{y_h} )\label{CV1}
\end{equation}
$\tau\equiv 1-T/T_C$ is the reduced temperature, the index $\nu$ is defined by the critical behavior of $\xi$, $\xi\propto|\tau|^{-\nu}$, and $C_0$ is a subtraction constant due to additive renormalization~\cite{BREZIN82}.
Eq.~\ref{CV1} is valid independent of the knowledge of the order parameter.
It must be stressed that the order of the transition can in priciple be established regardless of any prejudice by looking at the behavior of the specific heat, which is therefore be the quantity of reference.

If the order parameter is known the following scaling law is expected for its susceptibility
\begin{equation}
\chi = L_s^{\gamma/\nu}\Phi_\chi(\tau L_s^{1/\nu}, am_q L_s^{y_h} )\label{CHI1}
\end{equation}
$\nu$, $\alpha$, $\gamma$ identify the order and universality class of the transition.
A good order parameter must obey Eq.~\ref{CHI1}.
\begin{figure*}[hbt!]
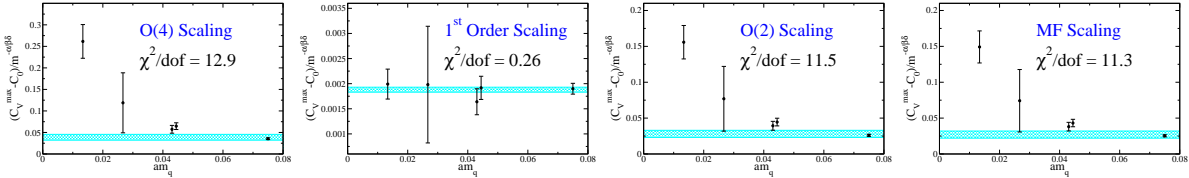

\includegraphics*[height=0.31\columnwidth]{Cv_max_O4.eps}
\includegraphics*[height=0.31\columnwidth]{Cv_max_1st.eps}
\includegraphics*[height=0.31\columnwidth]{Cv_max_O2.eps}
\includegraphics*[height=0.31\columnwidth]{Cv_max_mf.eps}
\vspace{-30pt}
\caption{Peak of the specific heat divided by the appropriate power of the mass (Eq.~\ref{CV}) to give a constant, for different scaling hypotheses.}\label{SH}
\end{figure*}

On the lattice $1/T = L_t a(\beta, m)$, so that 
\begin{equation}
\tau=1 - \frac{a(\beta_c, 0)}{a(\beta,m)} \approx C \left[ \beta_c - \beta + k m\label{REDT} \right]
\end{equation}
for sufficiently small masses while higher order terms in $m$ can enter at higher masses.
Here
\begin{equation}
C\equiv\left[\frac{\partial\ln a(\beta,m)}{\partial\beta}\right]_{(\beta,0)}\ ,\ k\equiv\left[\frac{\frac{\partial\ln a}{\partial m}}{\frac{\partial\ln a}{\partial\beta}}\right]_{(\beta,0)}
\end{equation}
In the quenched case $k=0$ and $\tau\propto\beta_c-\beta$, as usually assumed.
In the existing literature for $N_f=2$ the $m$ dependence in Eq.~\ref{REDT} is usually neglected, and $\tau$ is set proportional to $\beta_c-\beta$.

Our strategy has been to assume the $O(4)$, $O(2)$ exponent $y_h=2.49$ (see Table~\ref{CRITEXP}) and to run by choosing $am_q$, $L_s$ such that $am_q L_s^{y_h}=const$.
\begin{table}[b!]
\caption{Critical exponents.}\label{CRITEXP}
\begin{tabular}{|c|c|c|c|c|c|}
\hline & $y_t$ & $y_h$ & $\nu$ & $\alpha$ & $\gamma$\\
\hline $O(4)$ & 1.34 & 2.49 & 0.75 & -0.23 & 1.48\\
\hline $O(2)$ & 1.49 & 2.49 & 0.67 & -0.01 & 1.33\\
\hline $MF$ & $3/2$ & $9/4$ & $2/3$ & 0 & 1\\
\hline $1^{st} Order$ & 3 & 3 & $1/3$ & 1 & 1\\
\hline
\end{tabular}
\end{table}
The problem has then only one relevant scale and $\nu$, $\alpha$, $\gamma$ can be determined: if $O(4)$ or $O(2)$ is the correct universality class the peaks should scale as $(C_V-C_0)^{peak}\propto L_s^{\alpha/\nu}$, $\chi^{peak}\propto L_s^{\gamma/\nu}$.
Moreover, by analiticity arguments~\cite{karsch} in the $m\rightarrow 0$ limit one expects, if the transition is second order, that
\begin{eqnarray}
C_V - C_0 &=& (am_q)^{-\alpha/(\nu y_h)} \Phi_C'(\tau L_s^{1/\nu})\label{CV2} \\
\chi_m - \chi_0 &=& (am_q)^{-\gamma/(\nu y_h)} \Phi_\chi' (\tau L_s^{1/\nu})\label{CHI2}
\end{eqnarray}
The pseudocritical coupling should then obey the scaling law
\begin{equation}
\beta_c-\beta_{pc}+km \simeq k' / L_s^{1/\nu}
\end{equation}

In the case of a first order transition the analiticity arguments used in deriving the scaling Eqs.~\ref{CV2}-\ref{CHI2} are no longer valid, however they are expected to hold for small volumes compared to the critical volume.
For larger volumes the susceptibility are expected to increase with the volumes.

Section~\ref{NUMRESULT} contains the results of the analysis.

An additional susceptibility has been analyzed~\cite{RHO}, i.e. $\rho=\partial\ln\langle\mu\rangle /\partial\beta$, where $\langle\mu\rangle$ is an order parameter for confinement detecting dual superconductivity of the vacuum.

\section{Monte Carlo Simulations}

The lattice action used in Monte Carlo simulations was the standard staggered action.
Configurations were updated using the \textit{Hybrid R} algorithm and taking care of systematic errors due to finite molecular dynamics integration step size and finite conjugate gradient inversion residue. 
Two different sets of MC simulations were performed, called in the following \textit{Run1} and \textit{Run2}, having fixed for each set the value of $am_q L_s^{y_h}$ as explained above while we set $L_t=4$.
The parameters of our runs and the correspondig total number of trajectories collected are reported in Table~\ref{runpar}.
For each value of $am_q$ and $L_s$ a number ($10\div 20$) of simulations at different $\beta$ values were performed in order to inspect and to have under control the whole interesting critical region. 
The multi-histogram reweigthing technique was exploited to obtain information at intermediate $\beta$ values.
\begin{figure*}[t!]
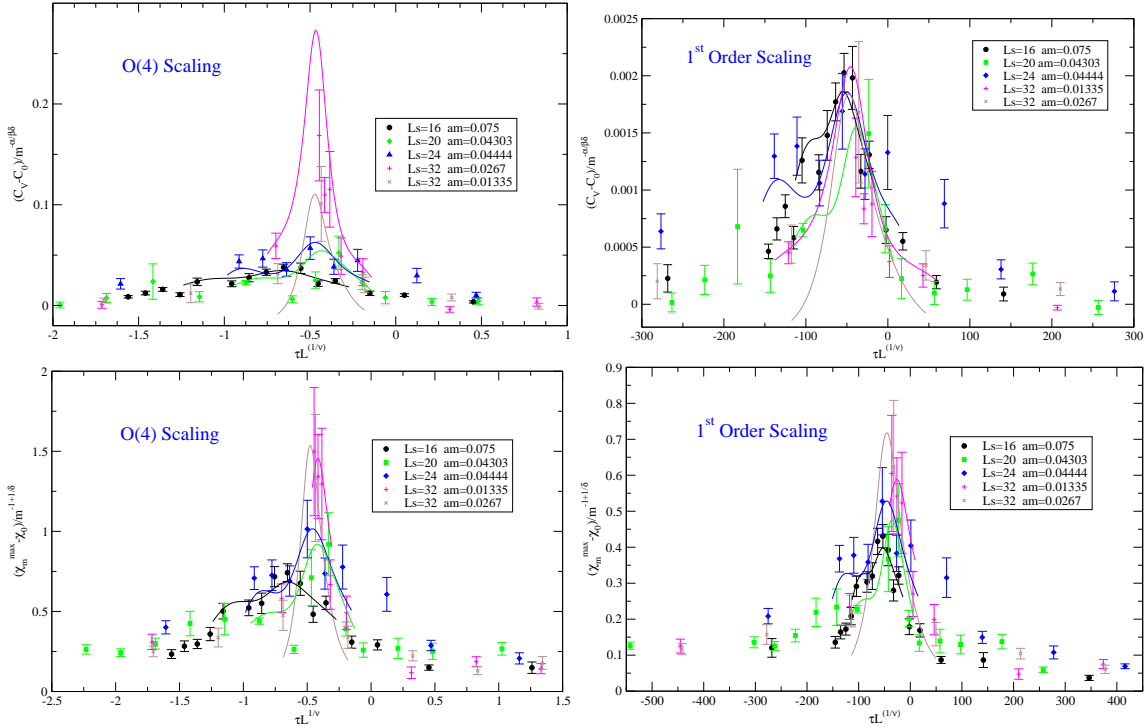

\includegraphics*[width=\columnwidth]{Cv_O4.eps}
\includegraphics*[width=\columnwidth]{Cv_1st.eps}\\
\includegraphics*[width=\columnwidth]{Chi_O4.eps}
\includegraphics*[width=\columnwidth]{Chi_1st.eps}
\vspace{-30pt}
\caption{Comparison of specific heat (top) and $\chi_m$ (bottom) peak scaling for $O(4)$ and $1^{st}$ order.}\label{SHCHISCA}
\end{figure*}
The following quantities were measured:
\begin{eqnarray}
C_V &=& \frac{1}{VT^2} \frac{\partial^2}{\partial \beta^2} \ln Z \longrightarrow \chi_{ij} , \chi_{ee} , \chi_{ie} \\
\chi_m &=& \frac{T}{V} \frac{\partial^2}{\partial m_q^2} \ln Z \approx V [\left<(\bar\psi\psi)^2\right>-\left<\bar\psi\psi\right>^2]\\
\chi_{ij} &=& V [\left<P_iP_j\right>-\left<P_i\right>\left<P_j\right>], \quad i,j = \sigma , \tau\\
\chi_{ee} &=& V [\left<(\bar\psi D_0 \psi)^2\right>-\left<\bar\psi D_0 \psi\right>^2] \\
\chi_{ie} &=& V [\left<P_i (\bar\psi D_0 \psi)\right>-\left<P_i\right>\left<\bar\psi D_0 \psi\right>]
\end{eqnarray}
where $V=L^3 N_t$ is the volume; $D_0$ is temporal component of the Dirac operator; $P_\sigma$, $P_\tau$ indicate the average spatial and temporal plaquette respectively; $C_V$ denotes the specific heat which is a function of $\chi_{ij}$, $\chi_{ee}$ and $\chi_{ie}$; $\chi_m$ is the susceptibility of the chiral condensate. Only the disconnected component of $\chi_m$ is considered in the present work since it gives the dominant contribution as it results from previous works and as it was also checked for a fraction of our MC simulations.
\begin{table}[b!]
\caption{Run parameters for the numerical simulations. \mbox{$L_s \cdot m_\pi$} varies in the range $[8.9 , 15.8]$}\label{runpar}
\begin{tabular}{|c|c|c|c|c|}
\hline & \multicolumn{2}{|c|}{$am_q$} & \multicolumn{2}{|c|}{\# Traj.} \\
\hline $L_s$ & Run1 & Run2 & Run1 & Run2 \\
\hline 12 & 0.153518 & 0.307036 & 22500 & 25000  \\
\hline 16 & 0.075 & 0.15 & 87700 & 131390  \\
\hline 20 & 0.04303 & 0.08606 & 14520 & 16100 \\
\hline 32 & 0.01335 & 0.0267 & 14500 & 15100 \\
\hline
\end{tabular}
\end{table}

\section{Numerical Results}\label{NUMRESULT}

Having fixed $am_q L_s^{y_h}=const$ with $y_h=2.49$ that is the value expected for $O(4)$ and $O(2)$ critical behavior, the following scaling formulas should hold (see Eqs.~\ref{CV1}-\ref{CHI1}):
\begin{eqnarray}
C_V(\tau, L) - C_0 = L^{\alpha/\nu} \Phi_C(\tau L_s^{1/\nu})\label{CV}\\
\chi_m(\tau, L) -\chi_0 = L^{\gamma/\nu} \Phi_\chi(\tau L_s^{1/\nu})\label{CHI}
\end{eqnarray}
In particular for the peaks of the susceptibilities we expect: $C_V^{peak}\propto L^{\alpha/\nu}$ and $\chi_m^{peak}\propto L^{\gamma/\nu}$. This behavior is only expected for the singular part of these quantities as the subtraction of constants $C_0$ and $\chi_0$ indicate in Eqs.~\ref{CV}-\ref{CHI}.
The measured peak values from data reweightening for the specific heat and $\chi_m$ for Run1 and Run2 are shown in Fig.~\ref{R12}. 
\begin{figure*}[t!]
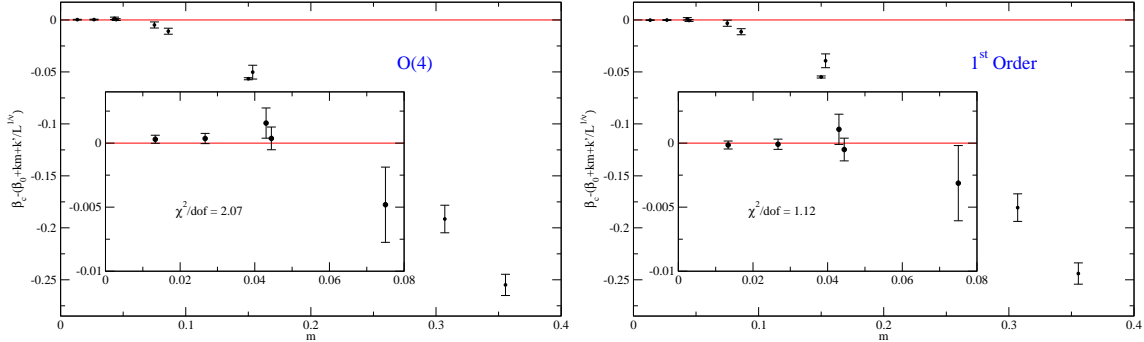

\includegraphics*[width=\columnwidth]{bc_O4.eps}
\includegraphics*[width=\columnwidth]{bc_1st.eps}
\vspace{-30pt}
\caption{Determination of the reduced temperature for $O(4)$ and $1^{st}$ order. The small $m$ region is enlarged in the inset picture.}\label{REDTFIT}
\end{figure*}
The $O(4)$ and $O(2)$ critical behavior is clearly in contradiction with the observed quantities. In particular $O(4)$ and $O(2)$ scaling predicts no singular behavior in the $L\rightarrow\infty$ limit for the specific heat as $\alpha$ is negative.
Also for $\chi_m$ the predicted exponents fail to reproduce lattice data.
In the whole study a non-improved action was used. One could then wonder whether this disagreement is a consequence of a poor convergence to the continuum limit. However it seems unlikely that an infrared qualitative feature of the theory, like the growth of the specific heat, can heavily rely on improvement of the action.

Wishing to test scaling behaviors other than the two previously considered, namely first order and mean field, we considered the scaling relations Eqs.~\ref{CV2}-\ref{CHI2}, valid in the limit $am_q\rightarrow 0$.
Smaller lattices and larger masses were discarded for this analysis in order to satisfy the constrains of above equations and an additional set of simulations with $L_s=24$ and $am_q=0.04444$ was added.
We repeated the analysis of the scaling of the peaks of susceptibilities and the results are shown in Fig.~\ref{SH} for the specific heat. $\chi_m$ shows a similar behavior. From the value of the $\chi^2/dof$ is easly seen that the only choice compatible with the data is a $1^{st}$ order scaling. Although the upper mass limit of validity of Eqs.~\ref{CV2}-\ref{CHI2} is not known \textit{a priori}, it should be noted that restricting the mass range of the fit worsen the agreement with $O(4)$, $O(2)$ and MF hypotheses while for first order it remains unchanged.
The scaling of susceptibilities at all $\beta$ can also be investigated. Fig.~\ref{SHCHISCA} shows the quality of the scaling for the two different working assumptions of $O(4)$ and $1^{st}$ order both for the specific heat and $\chi_m$. Even if it is difficult to make precise quantitative statements about the quality of these scalings, the first order case seems to give better results for both of the susceptibilities considered. This is clearly the case for the height of the susceptibilities as shown above, but it is also true for the width of these curves. The horizontal scale of these plots is given by the scaling variable $\tau L_s^{1/\nu}$ where $\tau$, the reduced temperature, depends on the unkwon value of $k$ (see Eq.~\ref{REDT}). 
The actual value of $k$ was extracted form the data with the fits shown in Fig.~\ref{REDTFIT}. Both $O(4)$ and first order critical exponent allow a good fit of $\tau$ so that it is not possible to discriminate between the two cases considering only pseudocritical couplings.

\begin{figure}[b!]
\includegraphics*[width=\columnwidth]{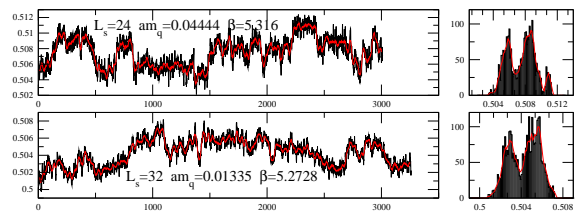}
\vspace{-30pt}
\caption{MC histories and observable histograms showing a double peak structure. Plaquette is actually shown and chiral condensate have a similar behavior.}\label{HIST}
\end{figure}
Having detected a signal of a first order phase transition, it is natural to look at time histories of observables to find signals of metastabilities. Such metastabilities should manistifest themselves as double peaks structures in the histograms of observable distributions. 
Indeed we found such structures for some of our time histories at largest volumes and smallest masses, as shown in Fig.~\ref{HIST}. Even though this is a preliminary indication of a first order phase transition it is not conclusive and we were not able to evidence this double peak structure consistently, in particular in the data taken at $am_q=0.0267$ with $L_s=32$, where it was expected to be present.

\section{Conclusions}

We have presented a systematic study addressing the fundamental question of the order of the chiral transition in two-flavor QCD.
A large numerical effort has been done with the aim of clarifing the nature of the chiral transition. Our MC simulations were run on $APEmille$ machines with a total computation cost of about 5~$TFlops\times month$.

The hypothesis of a second order phase transition belonging to the universality class of $O(4)$ or $O(2)$ has been carefully tested on the lattice by looking at the finite size scaling of the basic thermodynamic susceptibilities, first of all the specific heat and then the chiral condensate susceptibility.
To do that, we have fixed the scaling variable $am_qL_s^{y_h}$ in Eqs.~\ref{CV1}-\ref{CHI1} and we have studied the residual $L_s$ dependence.
Our results, presented in the first part of Sect.~\ref{NUMRESULT}, seem to exclude both the $O(4)$ and $O(2)$ universality classes.
In particular, the growth of the specific heat going towards small masses and large volumes is in constrast with the expected behavior of a $O(4)$ or $O(2)$ transition in which there should be no singular contribution to $C_V$.
The finite size scaling of the chiral condensate susceptibility is also incompatible with $O(4)$ and $O(2)$ universality class.

An extented analysis at small masses, presented at the end of Sect.~\ref{NUMRESULT}, clearly shows that a first order phase transition is instead consistent with lattice data while all other open choises are disfavored.
Together with the scaling of the specific heat and order parameter susceptibility, also the time hystories of observables show some signals of metastabilities, i.e. double peak structures in distributions, which, although not conclusive, are a qualitative indication of a first order phase transition.

We have used the standard staggered action and $L_t=4$. We plan to repeat the analysis with larger $L_t$ and improved action. However the growth of the peaks which excludes $O(4)$ and $O(2)$ is a typical infrared phenomenon, most likely independent on short range improvements.
We also plan to run keeping the scaling variable $am_qL_s^{y_h}$ constant under the assumption of a first order transition, to check directly if it is consistent.
The indication~\cite{RHO} coming from the dual superconductivity disorder parameter is also in the direction of a first order phase transition.

\end{document}